\pdfoutput=1
\RequirePackage{ifpdf}
\ifpdf 
\documentclass[pdftex]{sigma}
\else
\documentclass{sigma}
\fi

\numberwithin{equation}{section}
\numberwithin{theorem}{section}
\numberwithin{proposition}{section}
\numberwithin{lemma}{section}
\numberwithin{corollary}{section}
\numberwithin{definition}{section}
\numberwithin{note}{section}
\begin{document}

\allowdisplaybreaks

\renewcommand{\thefootnote}{$\star$}

\renewcommand{\PaperNumber}{026}

\FirstPageHeading

\ShortArticleName{A Quasi-Lie Schemes Approach to Second-Order Gambier Equations}

\ArticleName{A Quasi-Lie Schemes Approach\\
to Second-Order Gambier Equations\footnote{This paper is a~contribution to the Special Issue ``Symmetries of
Dif\/ferential Equations: Frames, Invariants and~Applications''.
The full collection is available at
\href{http://www.emis.de/journals/SIGMA/SDE2012.html}{http://www.emis.de/journals/SIGMA/SDE2012.html}}}

\Author{Jos\'e F.~CARI\~NENA~$^\dag$, Partha~GUHA~$^{\ddag}$ and Javier~DE LUCAS~$^{\S}$}

\AuthorNameForHeading{J.F.~Cari\~nena, P.~Guha and J.~de Lucas}

\Address{$^\dag$~Department of Theoretical Physics and IUMA, University of Zaragoza,
\\
\hphantom{$^\dag$}~Pedro Cerbuna 12, 50.009, Zaragoza, Spain}
\EmailD{\href{mailto:jfc@unizar.es}{jfc@unizar.es}}

\Address{$^{\ddag}$~S.N.~Bose National Centre for Basic Sciences, JD Block, Sector~III,
\\
\hphantom{$^\dag$}~Salt Lake, Kolkata - 700.098, India}
\EmailD{\href{mailto:partha@bose.res.in}{partha@bose.res.in}}

\Address{$^{\S}$~Faculty of Mathematics and Natural Sciences, Cardinal Stefan Wyszy\'nski University,
\\
\hphantom{$^\S$}~W\'oy-cickiego 1/3, 01-938, Warsaw, Poland}
\EmailD{\href{mailto:j.delucasaraujo@uksw.edu.pl}{j.delucasaraujo@uksw.edu.pl}}

\ArticleDates{Received September 26, 2012, in f\/inal form March 14, 2013; Published online March 26, 2013}

\Abstract{A {\it quasi-Lie scheme} is a~geometric structure that provides $t$-dependent changes of variables
transforming members of an associated family of systems of f\/irst-order dif\/ferential equations into members of the
same family.
In this note we introduce two quasi-Lie schemes for studying second-order Gambier equations in a~geometric way.
This allows us to study the transformation of these equations into simpler canonical forms, which solves a~gap in the
previous literature, and other relevant dif\/ferential equations, which leads to derive new constants of motion for
families of second-order Gambier equations.
Additionally, we describe general solutions of certain second-order Gambier equations in terms of particular solutions
of Riccati equations, linear systems, and $t$-dependent frequency harmonic oscillators.}


\Classification{34A26; 34A05; 34A34; 17B66; 53Z05}

\renewcommand{\thefootnote}{\arabic{footnote}} \setcounter{footnote}{0}

\vspace{-3mm}

\section{Introduction}
Apart from their inherent mathematical interest, dif\/ferential equations are important due to
their use in all branches of science~\cite{L68,Pi96}.
This strongly motivates their analysis as a~means to study the problems they model.
A remarkable approach to dif\/ferential equations is given by geometric methods~\cite{Olver}, which have resulted in
powerful techniques such as Lax pairs, Lie symmetries, and others~\cite{Ol92,OR87}.

A particular class of systems of ordinary dif\/ferential equations that have been drawing some attention in recent
years are the so-called {\it Lie systems}~\cite{AHW81,Dissertationes,HWA83,LS,SW84,PW}.
Lie systems form a~class of systems of f\/irst-order dif\/ferential equations possessing a~{\it superposition rule},
i.e.\ a~function that enables us to write the general solution of a~f\/irst-order system of dif\/ferential equations in
terms of a~generic collection of particular solutions and some constants to be related to initial
conditions~\cite{BM09II,CGM07}.

The theory of Lie systems furnishes many geometric methods for studying these
systems~\cite{BM09I,CGM00,CarRamGra,CMN,CarRam,Ru10,FLV10,PW}.
For instance, superposition rules can be employed to simplify the use of numerical techniques for solving
dif\/ferential equations~\cite{PW}, and the theory of reduction of Lie systems reduces the integration of Lie systems
on Lie groups to solving Lie systems on simple Lie groups~\cite{BM09I,CarRamGra}.

The classif\/ication of systems admitting a~superposition rule is due to Lie.
His result, the nowadays called Lie--Schef\/fers theorem, states that a~system admits a~superposition rule if and only
if it describes the integral curves of a~$t$-dependent vector f\/ield taking values in a~f\/inite-dimensional Lie
algebra of vector f\/ields.
The existence of such Lie algebras on $\mathbb{R}$ and $\mathbb{R}^2$ was analysed by Lie in his famous
work~\cite{Lie1880}.
More recently, the topic was revisited by Olver and coworkers~\cite{GKP92}, who clarif\/ied a~number of details that
were not properly described in the previous literature.

Despite their interesting properties, Lie systems have a~relevant drawback: there exist just a~few Lie systems of broad
interest~\cite{In72}.
Indeed, the Lie--Schef\/fers theorem and, more specif\/ically, the classif\/ication of f\/inite-dimensional Lie
algebras of vector f\/ields on low dimensional manifolds~\cite{GKP92,Lie1880} clearly show that that being a~Lie system
is the exception rather than the rule.
This has led to generalise the theory of Lie systems so as to tackle a~larger family of remarkable
systems~\cite{BecGagHusWin87,BecGagHusWin90,Dissertationes,In72}.
In particular, we henceforth focus on the so-called {\it quasi-Lie schemes}.
These recently devised structures~\cite{CGL10,CLL09Emd} have been found quite successful in investigating
transformation and integrability properties of dif\/ferential equations, e.g.\;Abel equations, dissipative
Milne--Pinney equations, second-order Riccati equations, and others~\cite{Dissertationes}.
In addition, the obtained results are useful so as to research on the physical and mathematical problems described
through these equations.

In this work, we study the second-order Gambier equations by means of the theory of quasi-Lie schemes.
We provide two new quasi-Lie schemes.
Their associated groups~\cite{CGL10} give rise to groups of $t$-dependent changes of variables, which are used to
transform second-order Gambier equations into another ones.
Such groups allow us to explain in a~geometric way the existence of certain transformations reducing a~quite general
subclass of second-order Gambier equations into simpler ones.
Our approach provides a~better understanding of a~result pointed out in~\cite{GRL98}.
As a~byproduct, we show that the procedure given in the latter work does not apply to every second-order Gambier
equation, which solves a~gap performed in there.

We provide conditions for second-order Gambier equations, written as f\/irst-order systems, to be mapped into Lie
systems via $t$-dependent changes of variables induced by our quasi-Lie schemes.
This is employed to determine families of Gambier equations which can be transformed into second-order Riccati
equations~\cite{GL99}, Kummer--Schwarz equations~\cite{Be98,Be97} and Milne--Pinney equations~\cite{Er80}.
These results are employed to derive, as far as we know, new constants of motion for certain second-order Gambier
equations.
Moreover, the description of their general solutions in terms of particular solutions of $t$-dependent frequency
harmonic oscillators, linear systems, or Riccati equations is provided~\cite{Dissertationes,GL12}.

\looseness=-1
The structure of our paper goes as follows.
Section~\ref{Section2} addresses the description of the fundamental notions to be employed throughout our work.
Section~\ref{Section3} describes a~new quasi-Lie scheme for studying second-order Gambier equations.
In Section~\ref{Section4} this quasi-Lie scheme is used to analyse the reduction of second-order Gambier equations to a~simpler
canonical form~\cite{GRL98}.
By using the theory of quasi-Lie systems, we determine in Section~\ref{Section5} a~family of second-order Gambier equations that can
be mapped into second-order Kummer--Schwarz equations.
The investigation of constants of motion for some members of the previous family is performed in Section~\ref{Section6}.
In Section~\ref{Section7} we describe the general solutions of a~family of second-order Gambier equations in terms of particular
solutions of other dif\/ferential equations.
We present a~second quasi-Lie scheme for investigating second-order Gambier equations in Section~\ref{Section8}, and conditions are
given to be able to transform these equations into second-order Riccati equations.
Those second-order Gambier equations that can be transformed into second-order Riccati equations are integrated in
Section~\ref{Section9}.
Finally, Section~\ref{Section10} is devoted to summarising our main results.

\section{Fundamentals}\label{Section2}

Let us survey the fundamental results to be used throughout the work (see~\cite{CGL10,CGL11,CGM00,CGM07} for details).
In general, we hereafter assume all objects to be smooth, real, and globally def\/ined on linear spaces.
This simplif\/ies our exposition and allows us to avoid tackling minor details.

Given the projection $\pi:(t,x)\in\mathbb{R}\times\mathbb{R}^n\mapsto x\in\mathbb{R}^n$ and the tangent bundle
projection $\tau:{\rm T}\mathbb{R}^n\rightarrow\mathbb{R}^n$, a~{\it $t$-dependent vector field} on $\mathbb{R}^n$ is
a~mapping $X:\mathbb{R}\times\mathbb{R}^n\rightarrow{\rm T}\mathbb{R}^n$ such that $\tau\circ X=\pi$.
This condition entails that every $t$-dependent vector f\/ield $X$ gives rise to a~family $\{X_t\}_{t\in\mathbb{R}}$ of
vector f\/ields $X_t:x\in\mathbb{R}^n\mapsto X(t,x)\in{\rm T}\mathbb{R}^n$ and vice versa.
We call {\it minimal Lie algebra} of $X$ the smallest real Lie algebra $V^X$ containing the vector f\/ields
$\{X_t\}_{t\in\mathbb{R}}$.
Given a~f\/inite-dimensional $\mathbb{R}$-linear space $V$ of vector f\/ields on $\mathbb{R}^n$, we write
$V(C^\infty(\mathbb{R}))$ for the $C^\infty(\mathbb{R})$-module of $t$-dependent vector f\/ields taking values in $V$.

An {\it integral curve} of $X$ is a~standard integral curve $\gamma:\mathbb{R}\rightarrow\mathbb{R}\times\mathbb{R}^n$
of its {\it suspension}, i.e.\
the vector f\/ield $\overline{X}=\partial/\partial t+X(t,x)$ on $\mathbb{R}\times\mathbb{R}^n$.
Note that the integral curves of $X$ of the form
$\gamma:t\in\mathbb{R}\rightarrow(t,x(t))\in\mathbb{R}\times\mathbb{R}^n$ are the solutions of the system
\begin{gather}
\label{Sys}
\frac{dx^i}{dt}=X^i(t,x),
\qquad
i=1,\ldots,n,
\end{gather}
the referred to as {\it associated system} of $X$.
Conversely, given such a~system, we can def\/ine a~$t$-dependent vector f\/ield on $\mathbb{R}^n$~\cite{CGL11}
\begin{gather*}
X(t,x)=\sum_{i=1}^nX^i(t,x)\frac{\partial}{\partial x^i}
\end{gather*}
whose integral curves of the form $(t,x(t))$ are the solutions to~\eqref{Sys}.
This justif\/ies to write $X$ for both a~$t$-dependent vector f\/ield and its associated system.

We call {\it generalised flow} a~map $g:(t,x)\in\mathbb{R}\times\mathbb{R}^n\mapsto g_t(x)\in\mathbb{R}^n$ such that
$g_0={\rm Id}_{\mathbb{R}^n}$.
Every $t$-dependent vector f\/ield $X$ can be associated with a~generalised f\/low $g$ satisfying that the general
solution of $X$ can be written in the form $x(t)=g_t(x_0)$ with $x_0\in\mathbb{R}^n$.
Conversely, every generalised f\/low def\/ines a~vector f\/ield by means of the expression~\cite{CGL10}
\begin{gather*}
X(t,x)=\frac{d}{ds}\bigg|_{s=t}g_s\circ g^{-1}_t(x).
\end{gather*}

Generalised f\/lows act on $t$-dependent vector f\/ields~\cite{CGM07}.
More precisely, given a~generalised f\/low $g$ and a~$t$-dependent vector f\/ield $X$, we can def\/ine a~unique
$t$-dependent vector f\/ield, $g_\bigstar X$, whose associated system has general solution $\bar x(t)=g_t(x(t))$, where
$x(t)$ is the general solution of $X$.
In other words, every $g$ induces a~$t$-dependent change of variables $\bar x(t)=g_t(x(t))$ transforming the system $X$
into $g_\bigstar X$.
Indeed, $g$ can be viewed as a~dif\/feomorphism
$\bar{g}:(t,x)\in\mathbb{R}\times\mathbb{R}^n\mapsto(t,g_t(x))\in\mathbb{R}\times\mathbb{R}^n$, and it can easily be
proved that $g_\bigstar X$ is the only $t$-dependent vector f\/ield such that $\overline{g_\bigstar
X}=\bar{g}_*\overline{X}$, where $\bar{g}_*$ is the standard action of the dif\/feomorphism $\bar g$ on vector f\/ields
(see~\cite{CGM07}).

Among all $t$-dependent vector f\/ields, we henceforth focus on those whose associated systems are Lie systems.
The characteristic property of Lie systems is to possess a~superposition rule~\cite{BM09II,CGM07,LS}.
A {\it superposition rule} for a~system $X$ on $\mathbb{R}^n$ is a~map
$\Phi:(u_{(1)},\ldots,u_{(m)};k_1,\ldots,k_n)\in(\mathbb{R}^{n})^m\times\mathbb{R}^n\mapsto\Phi(u_{(1)},
\ldots,u_{(m)};k_1,\ldots,k_n)\in\mathbb{R}^n$
allowing us to write its general solution $x(t)$ as
\begin{gather*}
x(t)=\Phi(x_{(1)}(t),\ldots,x_{(m)}(t);k_1,\ldots,k_n),
\end{gather*}
for a~generic family of particular solutions $x_{(1)}(t),\ldots,x_{(m)}(t)$ and a~set of constants $k_1,\ldots,k_n$ to
be related to initial conditions.

The celebrated Lie--Schef\/fers theorem~\cite[Theorem 44]{LS} states that a~system $X$ possesses a~superposition rule
if and only if it is a~$t$-dependent vector f\/ield taking values in a~f\/inite-dimensional real Lie algebra of vector
f\/ields, termed {\it Vessiot--Guldberg Lie algebra}~\cite{Ib09,RA00}.
In other words, $X$ is a~Lie system if and only if $V^X$ is f\/inite-dimensional~\cite{Dissertationes}.
This is indeed the main reason to def\/ine $V^X$~\cite{CLS12Ham}.

To illustrate the above notions, let us consider the Riccati equation~\cite{In56}
\begin{gather}
\label{Riccati}
\frac{dx}{dt}=b_1(t)+b_2(t)x+b_3(t)x^2,
\end{gather}
where $b_1(t)$, $b_2(t)$, $b_3(t)$ are arbitrary functions of time.
Its general solution, $x(t)$, can be obtained from an expression~\cite{Ko83,PW}
\begin{gather*}
x(t)=\Phi(x_{(1)}(t),x_{(2)}(t),x_{(3)}(t);k),
\end{gather*}
where $k$ is a~real number to be related to the initial conditions of every particular solution,
$x_{(1)}(t)$, $x_{(2)}(t)$, $x_{(3)}(t)$ are three dif\/ferent particular solutions of~\eqref{Riccati} and
$\Phi:\mathbb{R}^3\times\mathbb{R}\rightarrow\mathbb{R}$ is given by
\begin{gather*}
\Phi(u_{(1)},u_{(2)},u_{(3)};k)=
\frac{u_{(1)}(u_{(2)}-u_{(3)})-ku_{(2)}(u_{(3)}-u_{(1)})}{(u_{(2)}-u_{(3)})-k(u_{(3)}-u_{(1)})}.
\end{gather*}
That is, the Riccati equations admit a~superposition rule.
Therefore, from the Lie--Schef\/fers Theorem, we infer that the $t$-dependent vector f\/ield $X$ associated to
a~Riccati equation is such that $V^X$ is f\/inite-dimensional.
Indeed,
\begin{gather*}
X=\big(b_1(t)+b_2(t)x+b_3(t)x^2\big)\frac{\partial}{\partial x}.
\end{gather*}
Taking into account that
$
X_1= {\partial}/{\partial x}$,
$X_2=x {\partial}/{\partial x}$,
$
X_3=x^2 {\partial}/{\partial x}
$
span a~f\/inite-dimensional real Lie algebra $V$ of vector f\/ields and $X_t=b_1(t)X_1+b_2(t)X_2+b_3(t)X_3$, we obtain
that $\{X_t\}_{t\in\mathbb{R}}\subset V$ and $V^X$ becomes a~(f\/inite-dimensional) Lie subalgebra of $V$.

{\sloppy The Lie--Schef\/fers theorem shows that just some few f\/irst-order systems are Lie sys\-tems~\mbox{\cite{Dissertationes,In72}}.
For instance, this theorem implies that all Lie systems on the real line are, up to a~change of variables, a~particular
case of a~linear or Riccati equation~\cite{Ve93}.
Therefore, many other important dif\/ferential equations cannot be studied through Lie systems
(see~\cite{Dissertationes} for examples of this).
In order to treat non-Lie systems, new techniques generalising Lie systems need to be developed.
We here focus on the theory of quasi-Lie schemes~\cite{CGL10,CLL09Emd}.

}

\begin{definition}
 Let $W$, $V$ be f\/inite-dimensional real vector spaces of vector f\/ields on $\mathbb{R}^n$.
We say that they form a~{\it quasi-Lie scheme} $S(W,V)$ if:
\begin{itemize}\itemsep=0pt
\item $W$ is a~vector subspace of $V$.
\item $W$ is a~Lie algebra of vector f\/ields, i.e.\
$[W,W]\subset W$.
\item $W$ normalises $V$, i.e.\
$[W,V]\subset V$.
\end{itemize}
\end{definition}

Associated to each quasi-Lie scheme, we have the $C^\infty(\mathbb{R})$-modules $W(C^\infty(\mathbb{R}))$ and
$V(C^\infty(\mathbb{R}))$ of $t$-dependent vector f\/ields taking values in $W$ and $V$, respectively.
Now, from the Lie algebra~$W$, we def\/ine the group $\mathcal{G}(W)$ of generalised f\/lows of $t$-dependent vector
f\/ields taking values in~$W$, the so-called {\it group of the scheme}.
The relevance of this group is due to the following theorem~\cite{CGL10}.

\begin{theorem}
\label{Th:Main}
Given a~quasi-Lie scheme $S(W,V)$, every generalised flow of $\mathcal{G}(W)$ acts transforming elements of
$V(C^\infty(\mathbb{R}))$ into members of $V(C^\infty(\mathbb{R}))$.
\end{theorem}

In other words, the elements of the group of a~scheme provide $t$-dependent changes of variables that transform systems
of $V(C^\infty(\mathbb{R}))$ into systems of this family.
Roughly speaking, we can understand this group as a~generalisation of the $t$-independent symmetry group of a~system:
apart from transforming the initial system into itself, the transformations of the group of a~scheme also may transform
the initial system into one of the ``same type''.
For instance, given a~Lie system $X$ associated with a~Vessiot--Guldberg Lie algebra $V$, then $S(V,V)$ becomes
a~quasi-Lie scheme.
The group $\mathcal{G}(V)$ allows us to transform $X$ into a~Lie system with a~Vessiot--Guldberg Lie algebra $V$.
This can be employed, for example, to transform Riccati equations into Riccati equations that can be easily integrated,
giving rise to methods to integrate Riccati equations.

In order to illustrate previous notions, we now turn to proving that quasi-Lie schemes allow us to cope with Abel
equations of f\/irst-order and f\/irst kind~\cite{CLRAbel}, i.e.\
\begin{gather}
\label{Abel}
\frac{dx}{dt}=b_1(t)+b_2(t)x+b_3(t)x^2+b_4(t)x^3,
\end{gather}
with $b_1(t),\ldots,b_4(t)$ being arbitrary $t$-dependent functions.
Indeed, if we f\/ix $W=\langle\partial/\partial x,x\partial/\partial x\rangle$, it can be proved that $S(W,V)$ and
$V=\langle\partial/\partial x,x\partial/\partial x,x^2\partial/\partial x,x^3\partial/\partial x\rangle$ is a~quasi-Lie
scheme and $X\in V(C^\infty(\mathbb{R}))$ for every $X$ related to an Abel equation~\eqref{Abel}.
The elements of $\mathcal{G}(W)$ transform Abel equations into Abel equations and geometrically recover the usual
$t$-dependent changes of variables used to study these equations.
This was employed in~\cite{CLRAbel} to describe integrability properties of Abel equations.

Given a~quasi-Lie scheme $S(W,V)$, certain systems in $V(C^\infty(\mathbb{R}))$ can be mapped into Lie systems
admitting a~Vessiot--Guldberg Lie algebra contained in $V$.
This enables us to study the transformed system through techniques from the theory of Lie systems and, undoing the
performed transformation, to obtain properties of the initial system under study~\cite{CGL10}.
\begin{definition}
Let $S(W,V)$ be a~quasi-Lie scheme and $X$ a~$t$-dependent vector f\/ield in $V(C^\infty(\mathbb{R}))$, we say that $X$
is a~{\it quasi-Lie system with respect to $S(W,V)$} if there exists a~gene\-ra\-li\-sed f\/low $g\in\mathcal{G}(W)$ and
a~Lie algebra of vector f\/ields $V_0\subset V$ such that $g_\bigstar X\in V_0(C^\infty(\mathbb{R}))$.
\end{definition}

\section{A new quasi-Lie scheme\\ for investigating second-order Gambier equations}\label{Section3}

The {\it Gambier equation}~\cite{GRL98,GGCG11} can be described as the coupling of two Riccati equations in cascade,
which can be given in the following form
\begin{gather*}
\frac{dy}{dt}=-y^2+a_1y+a_2,
\\
\frac{dx}{dt}=a_0x^2+nyx+\sigma,
\end{gather*}
where $n$ is an integer, $\sigma$ is a~constant, which can be scaled to 1 unless it happens to be~$0$, and
$a_0$, $a_1$, $a_2$ are certain functions depending on time.
The precise form of the coef\/f\/icients of the Gambier equation is determined by singularity analysis, which leads to
some constraints on~$a_0$,~$a_1$ and~$a_2$~\cite{GRL98}.
For simplicity, we hereafter assume $a_0(0)\neq0$.
Nevertheless, all our results can easily be generalised for the case $a_0(0)=0$.

If $n\neq0$, we can eliminate~$y$ between the two equations above, which gives rise to the referred to as {\it
second-order Gambier equation}~\cite{GGCG11,LGR98,LGRW00,LR00}, i.e. {\samepage
\begin{gather}
\frac{d^2x}{dt^2}=
\frac{n-1}{xn}\left(\frac{dx}{dt}\right)^2+a_0\frac{(n+2)}{n}x\frac{dx}{dt}
+a_1\frac{dx}{dt}-\sigma\frac{(n-2)}{nx}\frac{dx}{dt}\nonumber
\\
\hphantom{\frac{d^2x}{dt^2}=}{}
-\frac{a_0^2}{n}x^3+\left(\frac{da_0}{dt}-a_0a_1\right)x^2+\left(a_2n-2a_0\frac{\sigma}{n}\right)x
-a_1\sigma-\frac{\sigma^2}{n x}.\label{GenGam}
\end{gather}  }

The importance of second-order Gambier equations is due to their relations to remarkable dif\/ferential equations such
as second-order Riccati equations~\cite{GL99,In56}, second-order Kummer--Schwarz equations~\cite{Be97,CGL11} and
Milne--Pinney equations~\cite{GGCG11}.
Additionally, by making appropriate limits in their coef\/f\/icients, Gambier equations describe all the linearisable
equations of the Painlev\'e--Gambier list~\cite{GGCG11}.
Several particular cases of these equations have also been studied in order to analyse discrete systems~\cite{LGRW00}.

Particular instances of~\eqref{GenGam} have already been investigated through the theory of Lie systems and quasi-Lie
schemes.
For instance, by f\/ixing $n=-2$, $\sigma=a_1=0$, and $a_0$ to be a constant, the second-order Gambier equation~\eqref{GenGam} becomes
a~second-order Kummer--Schwarz equation (KS2 equation)~\cite{Be97,CGL11}
\begin{gather}
\label{KS2}
\frac{d^2x}{dt^2}=\frac{3}{2x}\left(\frac{dx}{dt}\right)^2-2c_0x^3+2\omega(t)x,
\end{gather}
where we have written $c_0=-a^2_0/4$, with $c_0$ a~non-positive constant, and $\omega(t)=-a_2(t)$ so as to keep, for
simplicity in following procedures, the same notion as used in the literature, e.g.\
in~\cite{CGL11}.
The interest of KS2 equations is due to their relations to other dif\/ferential equations of physical and mathematical
interest~\cite{CGL11,Co94,GGCG11}.
For instance, for $x>0$ the change of variables $y=1/\sqrt{x}$ transforms KS2 equations into Milne--Pinney equations,
which frequently occur in cosmology~\cite{GGCG11}.
Meanwhile, the non-local transformation $dy/dt=x$ maps KS2 equations into a~particular type of third-order
Kummer--Schwarz equations, which are closely related to Schwarzian derivatives~\cite{Be98,CGL11,LG99}.
Additionally, KS2 equations can be related, through the addition of the new variable $dx/dt=v$, to a~Lie system
associated to a~Vessiot--Guldberg Lie algebra isomorphic to $\mathfrak{sl}(2,\mathbb{R})$, which gave rise to various
methods to study its properties and related problems~\cite{CGL11}.

If we now assume $n=1$ and $\sigma=0$ in~\eqref{GenGam}, it results
\begin{gather*}
\frac{d^2x}{dt^2}=\left(a_1+3a_0x\right)\frac{dx}{dt}-{a_0^2}x^3+\left(\frac{da_0}{dt}-a_0a_1\right)x^2+a_2x,
\end{gather*}
which is a~particular case of second-order Riccati equations~\cite{CRS05,GL99} that has been treated through the theory
of quasi-Lie schemes and Lie systems in several works~\cite{CL09SRicc,CLS12,GL12}.
Furthermore, equations of this type have been broadly investigated because of its appearance in the study of the
B\"acklund transformations for PDEs, their relation to physical problems, and the interest of the algebraic structure
of their Lie symmetries~\cite{BFL91,CRS05,GL99,KL09,LFB88}.

In view of the previous results, it is natural to wonder which kind of second-order Gambier equations can be studied
through the theory of quasi-Lie schemes.
To this end, let us build up a~quasi-Lie scheme for analysing these equations.

As usual, the introduction of the new variable $v\equiv dx/dt$ enables us to relate the second-order Gambier
equation~\eqref{GenGam} to the f\/irst-order system
\begin{gather*}
\frac{dx}{dt}=v,
\\
\frac{dv}{dt}=
\frac{(n-1)}{n}\frac{v^2}{x}+a_0\frac{(n+2)}{n}x v+a_1v-\sigma\frac{(n-2)}{n}\frac{v}{x}-\frac{a_0^2}{n}x^3
\\
\phantom{\frac{dv}{dt}=}{}
+\left(\frac{da_0}{dt}-a_0a_1\right)x^2+\left(a_2n-2a_0\frac{\sigma}{n}\right)x-a_1\sigma-\frac{\sigma^2}{n x},
\end{gather*}
which is associated to the $t$-dependent vector f\/ield on ${\rm T}\mathbb{R}_0$, with
$\mathbb{R}_0\equiv\mathbb{R}\setminus \{0\}$, given by
\begin{gather*}
X=
v\frac{\partial}{\partial x}+\left[\frac{(n-1)}{n}\frac{v^2}{x}
+a_0\frac{(n+2)}{n}x v+a_1v-\sigma\frac{(n-2)}{n}\frac{v}{x}-\frac{a_0^2}{n}x^3\right.
\\
\left.
\phantom{X=}{}
+\left(\frac{da_0}{dt}-a_0a_1\right)x^2+\left(a_2n-2a_0\frac{\sigma}{n}\right)x
-a_1\sigma-\frac{\sigma^2}{n x}\right]\frac{\partial}{\partial v},
\end{gather*}
termed henceforth the {\it Gambier vector field}.
To obtain a~quasi-Lie scheme for studying the above equations, we need to f\/ind a~f\/inite-dimensional
$\mathbb{R}$-linear space~$V_G$ such that $X\in V_G(C^\infty(\mathbb{R}))$ for all~$a_0$, $a_1$, $a_2$, $\sigma$ and $n$.
Observe that~$X$ can be cast in the form
\begin{gather}
\label{Dec}
X=\sum_{\alpha=1}^{10}b_\alpha\left(a_0,\frac{da_0}{dt},a_1,a_2,\sigma,n\right)Y_\alpha,
\end{gather}
with $b_1,\ldots,b_{10}$ being certain $t$-dependent functions whose form depends on the functions $a_0$,
$da_0/dt$, $a_1$, $a_2$ and the constants $\sigma$ and $n$.
More specif\/ically, these functions read
\begin{gather*}
b_1=1,
\qquad
b_2=\frac{n-1}{n},
\qquad
b_3=a_0\frac{n+2}{n},
\qquad
b_4=a_1,
\qquad
b_5=-\sigma\frac{n-2}{n},
\\
b_6=-\frac{a_0^2}{n},
\qquad
b_7=\frac{da_0}{dt}-a_0a_1,
\qquad
b_8=a_2n-2a_0\frac{\sigma}{n},
\qquad
b_9=-a_1\sigma,
\qquad
b_{10}=-\frac{\sigma^2}n
\end{gather*}
and
\begin{gather*}
Y_1=v\frac{\partial}{\partial x},
\qquad
Y_2=\frac{v^2}{x}\frac{\partial}{\partial v},
\qquad
Y_3=xv\frac{\partial}{\partial v},
\qquad
Y_4=v\frac{\partial}{\partial v},
\qquad
Y_5=\frac{v}{x}\frac{\partial}{\partial v},
\\
Y_6=x^3\frac{\partial}{\partial v},
\qquad
Y_7=x^2\frac{\partial}{\partial v},
\qquad
Y_8=x\frac{\partial}{\partial v},
\qquad
Y_9=\frac{\partial}{\partial v},
\qquad
Y_{10}=\frac{1}{x}\frac{\partial}{\partial v}.
\end{gather*}
For convenience, we further def\/ine the vector f\/ield
\begin{gather*}
Y_{11}=x\frac{\partial}{\partial x},
\end{gather*}
which, although does not appear in the decomposition~\eqref{Dec}, will shortly become useful so as to describe the
properties of Gambier vector f\/ields.

In view of~\eqref{Dec}, it easily follows that we can choose $V_G$ to be the space spanned by $Y_1,\ldots,Y_{11}$.
It is interesting to note that the linear space $V_G$ is not a~Lie algebra as $[Y_3,Y_6]$ does not belong to $V_G$.
Moreover, as
\begin{gather*}
{\rm ad}_{Y_3}^jY_6\equiv  \ \stackrel{\text{$j$-times}}{\overbrace{[Y_3,[Y_3,[\ldots,[Y_3}},Y_6]\ldots]]]=
(-1)^jx^{j+3}\frac{\partial}{\partial v},
\qquad
j\in\mathbb{N},
\end{gather*}
there is no f\/inite-dimensional real Lie algebra $\widehat V\supset V_G$ such that $X\in\widehat
V(C^\infty(\mathbb{R}))$.
Hence, $X$ is not in general a~Lie system, which suggests us to use quasi-Lie schemes to investigate it.

To determine a~quasi-Lie scheme involving $V_G$, we must f\/ind a~real f\/inite-dimensional Lie algebra $W_G\subset
V_G$ such that $[W_G,V_G]\subset V_G$.
In view of Table~\ref{Ta}, we can do so by setting $W_G=\langle Y_4,Y_8,Y_{11}\rangle$, which is a~solvable three-dimensional
Lie algebra.
In fact,
\begin{gather*}
[Y_4,Y_8]=-Y_8,
\qquad
[Y_4,Y_{11}]=0,
\qquad
[Y_8,Y_{11}]=-Y_8.
\end{gather*}
In other words, we have proved the following proposition providing a~new quasi-Lie scheme to study Gambier vector
f\/ields and, as shown posteriorly, second-order Gambier equations.

\begin{table}\centering
\caption{Lie brackets $[Y_i,Y_j]$ with $i=4,8,11$ and $j=1,\ldots,11$.}\label{Ta}
\vspace{1mm}

\begin{tabular}
{ | c | c | c | c | c | c | c | c | c | c | c | c |}
\hline
$[\cdot,\cdot]$ & $Y_1$ & $Y_2$ & $Y_3$ & $Y_4$ & $Y_5$ & $Y_6$ & $Y_7$ &$Y_8$ & $Y_9$ & $Y_{10}$&$Y_{11}$
\\
\hline
$Y_4$ & $Y_1$ & $Y_2$ & $0$ & $0$ & $0$ & $-Y_6$& $-Y_7$ &$-Y_8$& $-Y_{9}$ &$-Y_{10}$&$0$
\\
\hline
$Y_{8}$ & $Y_{11}-Y_{4}$& $2Y_4$& $Y_7$ & $Y_8$ &$Y_9$ &$0$ &$0$ &$0$ & $0$ &$0$ &$-Y_8$
\\
\hline
$Y_{11}$ & $-Y_1$ & $-Y_2$& $Y_3$& $0$ &$-Y_5$ &$3Y_6$ &$2Y_7$ &$Y_8$ &$0$ &$-Y_{10}$&$0$
\\
\hline
\end{tabular}
\end{table}

\begin{proposition}
The spaces $V_G=\langle Y_1,\ldots,Y_{11}\rangle$ and $W_G=\langle Y_4,Y_8,Y_{11}\rangle$ form a~quasi-Lie scheme
$S(W_G,V_G)$ such that $X\in V_G(C^\infty(\mathbb{R}))$ for every Gambier vector field~$X$.
\end{proposition}

Recall that $Y_{11}$ is not necessary so that every Gambier vector f\/ield takes values in~$V_G$.
Hence, why it is convenient to add it to~$V_G$? One reason can be found in Table~\ref{Ta}.
If $V_G$ had not contained~$Y_{11}$, then $S(W_G,V_G)$ would have not been a~quasi-Lie scheme as $W_G\nsubseteq V_G$.
In addition, $Y_8$ could not belong to $W_G$ neither, as $[Y_8,Y_1]\in V_G$ provided $Y_{11}-Y_4\in V_G$.
Hence, including $Y_{11}$ in $V_G$ allows us to choose a~larger $W_G\subset V_G$.
In turn this gives rise to a~larger group~$\mathcal{G}(W_G)$, which will be of great use in following sections.

\section{Transformation properties of second-order Gambier equations}\label{Section4}

Remind that Theorem~\ref{Th:Main} states that the $t$-dependent changes of variables associated to the elements of the
group $\mathcal{G}(W)$ of a~quasi-Lie scheme $S(W,V)$ establish bijections among the $t$-dependent vector f\/ields
taking values in~$V$.
This may be of great use so as to transform their associated systems into simplif\/ied forms, e.g.\
in the case of Abel equations~\cite{CLRAbel}.
We next show how this can be done for studying the transformation of second-order Gambier equations into simpler ones
whose corresponding coef\/f\/icients~$a_1$ vanish~\cite{GRL98}.
This retrives known results from a~geometrical viewpoint and shows that certain second-order Gambier equations cannot
be transformed into simpler ones, solving a~small gap performed in~\cite{GRL98}.

As the vector f\/ields in $W_G$ span a~f\/inite-dimensional real Lie algebra of vector f\/ields on ${\rm
T}\mathbb{R}_0$, there exists a~local Lie group action $\varphi:G\times{\rm T}\mathbb{R}_0\rightarrow{\rm
T}\mathbb{R}_0$ whose fundamental vector f\/ields are the elements of~$W_G$.
By integrating the vector f\/ields of~$W_G$ (see~\cite{SIGMA} for details), the action can easily be written as
\begin{gather*}
\varphi\left(g,\left(
\begin{matrix}x
\\
v
\\
\end{matrix}
\right)\right)=\left(
\begin{matrix}\alpha x
\\
\gamma x+\delta v
\\
\end{matrix}
\right),
\qquad \text{where}\qquad
g\in T_d\equiv\left\{\left(
\begin{matrix}\alpha&0
\\
\gamma&\delta
\\
\end{matrix}
\right)\bigg|\,\alpha,\delta\in\mathbb{R}_+,\,\gamma\in\mathbb{R}\right\}.
\end{gather*}
The theory of Lie systems~\cite{Dissertationes,CGM00} states that the solutions of a~system associated to
a~$t$-dependent vector f\/ield taking values in the real Lie algebra $W_G$ are of the form
$(x(t),v(t))=\varphi(h(t),(x_0,v_0))$, with $h(t)$ being a~curve in $T_d$ with $h(0)=e$.
Therefore, every $g\in\mathcal{G}(W_G)$ can be written as $g_t(\cdot)=\varphi(h(t),\cdot)$ for a~certain curve~$h(t)$
in $G$ with $h(0)=e$.
Conversely, given a~curve $h(t)$ in $T_d$ with $h(0)=e$, the curve $(x(t),v(t))=\varphi(h(t),(x_0,v_0))$ is the general
solution of a~system of $W_G(C^\infty(\mathbb{R}))$, which leads to a~generalised f\/low
$g_t(\cdot)=\varphi(h(t),\cdot)$ of $\mathcal{G}(W_G)$~\cite{Dissertationes,CGM00}.
Hence, the elements of $\mathcal{G}(W_G)$ are generalised f\/lows of the form
\begin{gather*}
g^{h(t)}(t,x,v)=\varphi(h(t),x,v),
\end{gather*}
for $h(t)$ any curve in $T_d$ with $h(0)=e$.
Observe that every $h(t)$ is a~matrix of the form
\begin{gather*}
h(t)=\left(
\begin{matrix}\alpha(t)&0
\\
\gamma(t)&\delta(t)
\\
\end{matrix}
\right)
\end{gather*}
where $\alpha(t)$, $\delta(t)$ and $\gamma(t)$ are $t$-dependent functions such that $\alpha(0)=\delta(0)=1$ and
$\gamma(0)=0$, because $h(0)=e$, and $\alpha(t)>0$, $\delta(t)>0$ as $h(t)\in T_d$ for every $t\in\mathbb{R}$.
Hence, every element of~$\mathcal{G}(W_G)$ is of the form
\begin{gather}
\label{EsTr}
g^{\alpha(t),\gamma(t),\delta(t)}(t,x,v)\equiv g^{h(t)}(t,x,v)=(t,\alpha(t)x,\gamma(t)x+\delta(t)v).
\end{gather}

Theorem~\ref{Th:Main} implies that for every $g\in\mathcal{G}(W_G)$ and Gambier vector f\/ield $X\in
V_G(C^\infty(\mathbb{R}))$, we have $g_\bigstar X\in V_G(C^\infty(\mathbb{R}))$.
More specif\/ically, a~long but straightforward calculation shows that
\begin{gather}
\label{Decomp}
g_\bigstar X=\sum_{\alpha=1}^{11}\bar{b}_\alpha(t)Y_\alpha,
\end{gather}
where the functions $\bar{b}_\alpha=\bar{b}_\alpha(t)$, with $\alpha=1,\ldots,11$, are
\begin{gather}
\bar{b}_1=\frac{\alpha}{\delta},
\qquad
\bar{b}_2=\frac{n-1}{n}\frac{\alpha}{\delta},
\qquad
\bar{b}_3=\frac{a_0(n+2)}{n\alpha},
\qquad
\bar{b}_4=a_1+\frac1{\delta}\frac{d\delta}{dt}+\frac{(2-n)\gamma}{n\delta},
\nonumber
\\
\bar{b}_5=\frac{(2-n)\sigma\alpha}{n},
\qquad
\bar{b}_6=-\frac{a_0^2\delta}{n\alpha^3},
\qquad
\bar{b}_7=\frac{1}{\alpha^2}\left(\delta\frac{da_0}{dt}-a_0a_1\delta-\frac{n+2}{n}a_0\gamma\right),
\nonumber
\\
\bar{b}_8=
\frac{\delta}{\alpha}\left(n a_2-\frac{2\sigma}{n}a_0-\frac{\gamma}{\delta}a_1-\frac{\gamma^2}{n\delta^2}
-\frac{\gamma}{\delta^2}\frac{d\delta}{dt}+\frac{1}{\delta}\frac{d\gamma}{dt}\right),
\nonumber
\\
\bar{b}_9=-\sigma\left(a_1\delta+\frac{(2-n)\gamma}{n}\right),
\qquad
\bar{b}_{10}=-\frac{\sigma^2\alpha\delta}{n},
\qquad
\bar{b}_{11}=\frac{1}{\alpha}\frac{d\alpha}{dt}-\frac{\gamma}{\delta}.
\label{NewCoef}
\end{gather}
Let us use this so as to transform the initial Gambier vector f\/ield into another one that is related, up to
a~$t$-reparametrisation $\tau=\tau(t)$, to a~second-order Gambier equation with $\tau$-dependent coef\/f\/icients $\bar
a_0$, $\bar a_1$, $\bar a_2$, a~constant $\bar\sigma$ and an integer number $\bar{n}$.
Additionally, we impose $\bar a_1=0$ in order to reduce our initial f\/irst-order Gambier equation into a~simpler one.
In this way, we have
\begin{gather}
g_\bigstar X=
\xi(t)\left[Y_1+\frac{\bar n-1}{\bar n}Y_2+\bar{a}_0\frac{\bar n+2}{\bar n}Y_3-\sigma\frac{\bar n-2}{\bar n}Y_5\right.
\nonumber
\\
\left.
\phantom{g_\bigstar X=}{}
-\frac{\bar{a}_0^2}{\bar n}Y_6+\frac{d\bar a_0}{d\tau}Y_7
+\left(\bar a_2\bar n-2\bar a_0\frac{\bar\sigma}{\bar n}\right)Y_8-\frac{\bar{\sigma}^2}{\bar n}Y_{10}\right],
\label{FinalForm}
\end{gather}
for a~certain non-vanishing function $\xi(t)=d\tau/dt$.
Therefore, $\bar{b}_4=\bar{b}_{9}=\bar{b}_{11}=0$, i.e.\
\begin{gather}
a_1+\dfrac1{\delta}\dfrac{d\delta}{dt}+\dfrac{(2-n)\gamma}{n\delta}=0,
\label{eq1}
\\
\sigma\left(a_1\delta+\dfrac{(2-n)\gamma}{n}\right)=0,
\label{eq2}
\\
\dfrac1{\alpha}\dfrac{d\alpha}{dt}-\dfrac{\gamma}{\delta}=0.
\label{eq3}
\end{gather}
As we want our method to work for all values of $\sigma$, e.g.
$\sigma\neq0$, equation~\eqref{eq2} implies
\begin{gather}
\label{eq2b}
a_1\delta+\frac{(2-n)\gamma}n=0.
\end{gather}
As $\delta>0$, the above equation involves that $a_1=0$ for $n=2$.
In other words, we cannot transform a~Gambier vector f\/ield with $a_1\neq0$ into a~new one with $\bar{a}_1=0$ through
our methods if $n=2$ and $\sigma\neq0$.
In view of this, let us assume that $n\neq2$.

From~\eqref{eq1} and~\eqref{eq2b}, and using again that $\delta>0$, we infer that $d\delta/dt=0$.
As $\delta(0)=1$, then $\delta=1$.
Plugging the value of $\delta$ into~\eqref{eq3} and~\eqref{eq2b}, we obtain
\begin{gather*}
\frac{1}{\alpha}\frac{d\alpha}{dt}=\frac{na_1}{n-2}=\gamma\quad \Longleftrightarrow \quad \alpha=
\exp\left(\int^t_0\frac{na_1}{n-2}dt'\right),
\qquad
\gamma=\frac{na_1}{n-2},
\end{gather*}
which f\/ixes the form of $g$ mapping a~system $X$ into~\eqref{FinalForm}.
Bearing previous results in mind, we see that the non-vanishing $t$-dependent coef\/f\/icients~\eqref{NewCoef} become
\begin{gather*}
\bar{b}_1={\alpha},
\qquad
\bar{b}_2=\frac{n-1}{n}{\alpha},
\qquad
\bar{b}_3=\frac{a_0(n+2)}{n\alpha},
\qquad
\bar{b}_5=\frac{(2-n)\sigma\alpha}{n},
\qquad
\bar{b}_6=-\frac{a_0^2}{n\alpha^3},\\
\bar{b}_7=\frac{d}{dt}\left(\frac{a_0}{\alpha^2}\right),
\qquad
\bar{b}_8=
\frac{1}{\alpha}\left(n a_2-\frac{2\sigma a_0}{n}-\frac{n(n-1)}{(n-2)^2}{a_1^2}+\frac{n}{n-2}{\frac{da_1}{dt}}\right),
\qquad
\bar{b}_{10}=-\frac{\sigma^2\alpha}{n}.
\end{gather*}
Comparing this and~\eqref{FinalForm}, we see that to transform the initial f\/irst-order Gambier equation into a~new
one through a~$t$-dependent change of variables~\eqref{EsTr} and a~$t$-reparametrisation $d\tau=\alpha dt$ requires
$\xi=\alpha$.
The resulting system reads
\begin{gather*}
\frac{d\bar x}{d\tau}=\bar v,
\\
\frac{d\bar v}{d\tau}=
\frac{(n-1)}{n}\frac{\bar{v}^2}{\bar x}+a_0\frac{(n+2)}{\alpha^2n}\bar x\bar v-\sigma\frac{(n-2)}{n}\frac{\bar v}{\bar x}
-\frac{a_0^2}{\alpha^4n}\bar{x}^3
\\
\phantom{\frac{d\bar v}{d\tau}=}
+\frac{d\big(a_0/\alpha^2\big)}{d\tau}\bar{x}^2
+\left(\bar{a}_2n-2{a}_0\alpha^{-2}\frac{\sigma}{n}\right)\bar x-\frac{\sigma^2}{n\bar{x}},
\end{gather*}
where
\begin{gather*}
\bar a_2=\frac{1}{\alpha^2}\left(a_2-\frac{(n-1)}{(n-2)^2}a_1^2+\frac{1}{n-2}\frac{da_1}{dt}\right).
\end{gather*}
Therefore, redef\/ining $\bar n=n$, $\bar a_0=a_0/\alpha^2$ and $\bar\sigma=\sigma$, we obtain that the above system is
associated to a~second-order Gambier equation with $\bar a_1=0$.
Meanwhile, as $g$ induces a~$t$-dependent change of variables given by
\begin{gather*}
\bar x = \alpha x,
\qquad
\bar v =  \frac{d\alpha}{d\tau}x +v,
\end{gather*}
we see that this $t$-dependent change of variables can be viewed as a~$t$-dependent change of variables $\bar
x=\alpha x$ along with a~$t$-reparametrisation $t\rightarrow\tau$ such that $\bar v=d\bar x/d\tau$.
Indeed,
\begin{gather*}
\bar x=\alpha x\quad\Longrightarrow\quad\frac{d\bar x}{d\tau}=\frac{d\alpha}{d\tau}x+\alpha\frac{dx}{d\tau}=
\frac{d\alpha}{d\tau}x+v=\bar v.
\end{gather*}
Furthermore, these transformations map the initial second-order Gambier equation with $n\neq2$ into a~new one with
$\bar a_1=0$, i.e.\
depending only on two functions $\bar a_0$ and $\bar a_2$ and the constants $\sigma$ and $n$.
We can therefore formulate the following result.
\begin{proposition}
\label{Prop:red}
Every second-order Gambier equation~\eqref{GenGam} with $n\neq2$ can be transformed via a~$t$-dependent change of
variables $\bar x=\alpha(t)x$ and a~$t$-reparametrisation $\tau=\tau(t)$, with $d\tau=\alpha\,dt$, $\alpha(0)=1$ and
\begin{gather*}
\frac{1}{\alpha}\frac{d\alpha}{dt}=\frac n{n-2} a_1,
\end{gather*}
into a~second-order Gambier equation whose $a_1$ vanishes and $n$, $\sigma$ remain the same.
\end{proposition}

In the above proposition, we excluded second-order Gambier equations with $n=2$ as we noticed that in this case the
proof of this proposition does not hold: we cannot transform an initial Gambier vector f\/ield with $\sigma\neq0$ and
$a_1\neq0$ into a~new one with $\bar a_1=0$.
Moreover, it is easy to see that the transformation provided in Proposition~\ref{Prop:red} does not exist for $n=2$ and
$a_1\neq0$.
This was not noticed in~\cite{GRL98}, where this transformation is wrongly claimed to transform any Gambier equation
into a~simpler one with $a_1=0$.
In view of this, we cannot neither ensure, as claimed in~\cite{GRL98}, that second-order Gambier equations are not
given in their simplest form.

\section{Quasi-Lie systems and Gambier equations}\label{Section5}

The theory of quasi-Lie schemes mainly provides information about
quasi-Lie systems, which can be mapped to Lie systems through one of the transformations of the group of a~quasi-Lie
scheme.
This allows us to employ the techniques of the theory of Lie systems so as to study the obtained Lie systems, and,
undoing the performed $t$-dependent change of variables, to describe properties of the initial system~\cite{CGL10}.

Motivated by the above, we determine and study the Gambier vector f\/ields $X\in V(C^\infty(\mathbb{R}))$ which are
quasi-Lie systems relative to $S(W_G,V_G)$.
This task relies in f\/inding triples $(g,X,V_0)$, with $g\in\mathcal{G}(W_G)$ and $V_0$ being a~real Lie algebra
included in $V_G$ in such a~way that $g_\bigstar X\in V_0(C^\infty(\mathbb{R}))$.

One of the key points to determine quasi-Lie systems is to f\/ind a~Lie algebra $V_0$.
In the case of Gambier vector f\/ields, this can readily be obtained by recalling that certain instances of
second-order Gambier equations, e.g.~\eqref{KS2}, are particular cases of KS2 equations.
By adding a~new variable $v\equiv dx/dt$ to~\eqref{KS2}, the resulting f\/irst-order system becomes a~Lie system
(see~\cite{CGL11}) related to a~three-dimensional Vessiot--Guldberg Lie algebra $V_0\subset V_G$ of vector f\/ields on
${\rm T}\mathbb{R}_0$ spanned by
\begin{gather}
\label{VFKS2}
X_1=2x\frac{\partial}{\partial v},
\qquad
X_2=x\frac{\partial}{\partial x}+2v\frac{\partial}{\partial v},
\qquad
X_3=v\frac{\partial}{\partial x}+\left(\frac32\frac{v^2}x-2c_0x^3\right)\frac{\partial}{\partial v},
\end{gather}
i.e.\
\begin{gather*}
X_1=2\,Y_8,
\qquad
X_2=Y_{11}+2\,Y_4,
\qquad
X_3=Y_1+\frac32Y_2-2\,c_0\,Y_6.
\end{gather*}
Consequently, it makes sense to look for Gambier vector f\/ields $X\in V(C^\infty(\mathbb{R}))$ that can be
transformed, via an element $g\in\mathcal{G}(W_G)$, into a~Lie system $g_\bigstar X\in V_0(C^\infty(\mathbb{R}))$, i.e.\
\begin{gather*}
g_\bigstar X=2f(t)Y_8+g(t)(Y_{11}+2Y_4)+h(t)\left(Y_1+\frac32Y_2-2c_0Y_6\right),
\end{gather*}
for certain $t$-dependent functions $f$, $g$ and $h$.
Comparing the expression of $g_\bigstar X$ given by~\eqref{NewCoef} and the above, we f\/ind that $g_\bigstar X\in
V_0(C^\infty(\mathbb{R}))$ if and only if
\begin{gather}
\label{rel0}
\bar{b}_3=\bar{b}_5=\bar{b}_7=\bar{b}_9=\bar{b}_{10}=0
\end{gather}
and
\begin{gather}
\label{rel2}
\bar{b}_4=2\bar{b}_{11},
\qquad
\bar{b}_2=\frac32\bar{b}_1,
\qquad
\bar{b}_6=-2c_0\bar{b}_1.
\end{gather}

From expressions~\eqref{NewCoef} and remembering that $\alpha>0$ and $\delta>0$, we see that condition $\bar{b}_{10}=0$
implies that $\sigma=0$.
This involves, along with~\eqref{NewCoef}, that $\bar{b}_5=\bar{b}_9=0$.
Meanwhile, from $\bar{b}_2=3\bar{b}_1/2$ we obtain $n=-2$, which in turn ensures that $\bar{b}_3=0$.
Bearing all this in mind, $\bar{b}_7=0$ reads
\begin{gather}
\label{int}
a_0a_1-\frac{da_0}{dt}=0.
\end{gather}
Above results impose restrictions on the form of the Gambier vector f\/ield $X$ to be able to be transformed into a~Lie
system possessing a~Vessiot--Guldberg Lie algebra $V_0$ via an element $g\in\mathcal{G}(W_G)$.
Let us show that the remaining conditions in~\eqref{rel0} and~\eqref{rel2} merely characterise the form of the
$t$-dependent change of variables $g$ and the coef\/f\/icient $c_0$ appearing in~\eqref{KS2}.

Conditions $2\bar{b}_{11}=\bar{b}_4$ and $\bar{b}_6=-2c_0\bar{b}_1$ read
\begin{gather}
\label{eq}
\frac{d}{dt}\log\frac{\alpha^2}{\delta}=a_1,
\qquad
a_0^2=-4c_0\frac{\alpha^4}{\delta^2}.
\end{gather}
Using the f\/irst condition above, the relation~\eqref{int}, and taking into account that $\alpha(0)=\delta(0)=1$ and
$\delta,\alpha>0$, we see that
\begin{gather*}
\frac{\frac{d}{dt}{(\alpha^2\delta^{-1})}}{\alpha^2\delta^{-1}}=
\frac{1}{a_0}\frac{da_0}{dt}\quad\Longrightarrow\quad\frac{\alpha^2}{\delta}=\frac{a_0}{a_0(0)}.
\end{gather*}
Using the second equality in~\eqref{eq}, we obtain
\begin{gather*}
4c_0=-a_0(0)^2,
\end{gather*}
which f\/ixes $c_0$ in terms of a~coef\/f\/icient of the initial second-order Gambier equation.

Concerning the $t$-dependent coef\/f\/icients $\bar{b}_1,\ldots,\bar{b}_{11}$, the non-vanishing ones under above
conditions, i.e.\
$\bar b_1$, $\bar b_2$, $\bar b_4$, $\bar b_6$, $\bar b_8$ and $\bar b_{11}$, can readily be obtained through
relations~\eqref{rel2} and
\begin{gather*}
\bar b_{1}=\frac{\alpha}{\delta},
\qquad
\bar b_8=
\frac{\delta}{\alpha}\left(-2a_2-\frac{\gamma}{\delta a_0}\frac{da_0}{dt}+\frac{\gamma^2}{2\delta^2}
-\frac{\gamma}{\delta^2}\frac{d\delta}{dt}+\frac{1}{\delta}\frac{d\gamma}{dt}\right),
\qquad
\bar b_{11}=\frac{1}{\alpha}\frac{d\alpha}{dt}-\frac{\gamma}{\delta}.
\end{gather*}
In other words, we have proved the following proposition.
\begin{proposition}
A Gambier vector field $X$ is a~quasi-Lie system relative to $S(W_G,V_G)$ that can be transformed into a~Lie system
taking values in a~Lie algebra of the form $V_0$ if and only if $a_0a_1=da_0/dt$, $n=-2$ and $\sigma=0$.
Under these conditions, the constant $c_0$ appearing in $X_3$ becomes $c_0=-a_0(0)^2/4$.
Additionally, a~transformation $g(\alpha(t),\gamma(t),\delta(t))\in\mathcal{G}(W_G)$ maps $X$ into $g_\bigstar X\in
V_0(C^\infty(\mathbb{R}))$ if and only if ${\alpha^2}/{\delta}={a_0}/{a_0(0)}$.
More specifically, $g_\bigstar X$ reads
\begin{gather}
\label{TranKS2}
g_\bigstar X=
\frac{\alpha}{\delta}X_3+\left(\frac{1}{\alpha}\frac{d\alpha}{dt}-\frac{\gamma}{\delta}\right)X_2
+\frac{\delta}{2\alpha}\left(-2a_2-\frac{\gamma}{\delta a_0}\frac{da_0}{dt}+\frac{\gamma^2}{2\delta^2}
-\frac{\gamma}{\delta^2}\frac{d\delta}{dt}+\frac{1}{\delta}\frac{d\gamma}{dt}\right)X_1.
\end{gather}
\end{proposition}

The above proposition allows us to determine the transformations $g\in\mathcal{G}(W_G)$ ensuring that a~Gambier vector
f\/ield and its related second-order Gambier equation can be mapped, maybe up to a~$t$-reparametrisation, into a~new
Gambier vector f\/ield related to a~KS2 equation.
Indeed, from~\eqref{KS2}, we see that to do so, we need to impose
\begin{gather*}
g_\bigstar X=\xi(t)(X_3+\omega(t)X_1),
\end{gather*}
for a~certain function $\omega(t)$ and a~function $\xi(t)$ such that $\xi(t)\neq0$ for all $t\in\mathbb{R}$.
Comparing~\eqref{TranKS2} with the above expression, we see that
\begin{gather}
\label{con4}
\frac{1}{\alpha}\frac{d\alpha}{dt}-\frac{\gamma}{\delta}=0,
\end{gather}
which, in view of the fact that $g^{\alpha(t),\gamma(t),\delta(t)}$ satisf\/ies
\begin{gather}
\label{con4-2}
\frac{\alpha^2}{\delta}=\frac{a_0}{a_0(0)},
\end{gather}
enables us to determine the searched transformations.
In fact, f\/ixed a~non-vanishing $t$-dependent function $\alpha$ with $\alpha(0)=1$, the above conditions determine the
values of $\delta$ and $\gamma$ of $g$.

Now, a~$t$-dependent reparametrisation
\begin{gather}
\label{tpar}
\tau=\int^t_0\frac\alpha\delta\;dt'
\end{gather}
transforms the system associated to $g_\bigstar X$ into a~new system related to $X_3+\omega(t(\tau))X_1$, where
$\omega$ is given by
\begin{gather}
\label{frequency}
\omega=
-\frac{\delta^2}{2\alpha^2}\left(2a_2+\frac{\gamma}{a_0\delta}\frac{da_0}{dt}-\frac{\gamma^2}{2\delta^2}
-\frac{1}{\delta}\frac{d\gamma}{dt}+\frac{\gamma}{\delta^2}\frac{d\delta}{dt}\right).
\end{gather}
\begin{proposition}
\label{Pro1}
Every Gambier vector field $X$ satisfying $a_0a_1=da_0/dt$, $n=-2$ and $\sigma=0$ is a~quasi-Lie system relative to
$S(W_G,V_G)$ that can be transformed into a~Lie system asso\-cia\-ted to $X_3+\omega(t)X_1$, with $c_0=-a_0(0)^2/4$ and
$\omega(t)$ given by~\eqref{frequency}, through a~transformation
$g^{\alpha(t),\gamma(t),\delta(t)}\in\mathcal{G}(W_G)$, whose coefficients are given by any solution
of~\eqref{con4} and~\eqref{con4-2}, and the $t$-dependent reparametrisation~\eqref{tpar}.
\end{proposition}

Note that the transformation $g^{\alpha(t),\gamma(t),\delta(t)}$ can be viewed as a~$t$-dependent change of variables
$\bar x=\alpha x$ and a~$t$-reparametrisation $d\tau=\alpha/\delta\, dt$, i.e.\
\begin{gather*}
\bar x=\alpha x,
\qquad
\bar v=\gamma x+\delta v,
\end{gather*}
and, in view of the f\/irst condition in~\eqref{con4}, we see that
\begin{gather*}
\frac{d\bar x}{d\tau}=\frac{d\alpha}{d\tau}x+\alpha\frac{dx}{d\tau}=\gamma x+\delta v=\bar v.
\end{gather*}
From this and Proposition~\ref{Pro1}, we obtain the following proposition about second-order Gambier equations.

\begin{corollary}
\label{Cor:trans}
Every second-order Gambier equation of the form
\begin{gather}
\label{GamQuasiLie}
\frac{d^2x}{dt^2}=
\frac{3}{2x}\left(\frac{dx}{dt}\right)^2+\frac{1}{a_0}\frac{d{a}_0}{dt}\frac{dx}{dt}+\frac{a_0^2}{2}x^3-2a_2x
\end{gather}
can be mapped into a~KS2 equation
\begin{gather}
\label{KS2three}
\frac{d^2\bar x}{d\tau^2}=
\frac{3}{2\bar x}\left(\frac{d\bar x}{d\tau}\right)^2+\frac12a_0(0)^2{\bar x}^3
-\frac{\delta^2}{\alpha^2}\left(2a_2+\frac{\gamma}{a_0\delta}\frac{da_0}{dt}-\frac{1}{2}\frac{\gamma^2}{\delta^2}
-\frac1{\delta}\frac{d\gamma}{dt}+\frac{\gamma}{\delta^2}\frac{d\delta}{dt}\right)\bar x
\end{gather}
through a~$t$-dependent change of variables $\bar x(t)=\alpha(t)x(t)$, where $\alpha(t)$ is any positive function with
$\alpha(0)=1$, and a~$t$-reparametrisation $\tau(t)$ with $d\tau=\alpha/\delta dt$, with $\delta$ and $\gamma$ being
determined from $\alpha$ by the relations
\begin{gather}
\label{con6}
\delta=\alpha^2\frac{a_0(0)}{a_0},
\qquad
\gamma=\frac{\alpha a_0(0)}{a_0}\frac{d\alpha}{dt}.
\end{gather}
\end{corollary}

\section{Constants of motion for second-order Gambier equations}\label{Section6}

In this section, we obtain constants of motion for second-order Gambier equations.
We do so by analysing the existence of $t$-independent constants of motion for systems $g_\bigstar X\in
V_0(C^\infty(\mathbb{R}))$, with $X$ being a~Gambier vector f\/ield.
By undoing the $t$-dependent change of variables $g$, this leads to determining constants of motion for a~Gambier
vector f\/ield $X$ and its corresponding second-order Gambier equation.

Let $F:{\rm T}\mathbb{R}_0\rightarrow\mathbb{R}$ be a~$t$-independent constant of motion for $g_\bigstar X$, we have
$(g_\bigstar X)_tF=0$ for all $t\in\mathbb{R}$.
This involves that $F$ is a~$t$-independent constant of motion for all successive Lie brackets of elements of
$\{(g_\bigstar X)_t\}_{t\in\mathbb{R}}$ as well as their linear combinations.
In other words, $F$ is a~common f\/irst-integral for the vector f\/ields $V^{g_\bigstar X}\subset V_0$.

When $\omega(t)$ is not a~constant, it can be verif\/ied that $V^{g_\bigstar X}=V_0$.
Thus, if $F$ is a~f\/irst-integral for all these vector f\/ields, it is so for all vector f\/ields contained in the
generalised distribution $\mathcal{D}_p=\langle(X_1)_p,(X_2)_p,(X_3)_p\rangle$, with $p\in{\rm T}\mathbb{R}_0$.
Hence, $dF_p\in\mathcal{D}_p^\circ$, i.e.\
$dF_p$ is incident to all vectors of~$\mathcal{D}_p$.
In this case, $\mathcal{D}_p={\rm T}_p{\rm T}\mathbb{R}_0$ for a~generic $p\in{\rm T}\mathbb{R}_0$, which implies that
$dF_p=0$ at almost every point.
Since $F$ is assumed to be dif\/ferentiable, we have that $F$ is constant on each connected component of ${\rm
T}\mathbb{R}_0$ and $g_\bigstar X$ has no non-trivial $t$-independent constant of motion.

If $\omega(t)=\lambda$ for a~real constant $\lambda$, then $\dim\mathcal{D}_p=1$ at a~generic point and it makes sense
to look for non-trivial $t$-dependent constants of motion for $g_\bigstar X$.
In view of~\eqref{frequency}, this condition implies
\begin{gather*}
\lambda=
-\frac{a_0(0)^2\alpha^2}{2a_0^2}\left[2a_2+\frac{1}{a_0}\frac{da_0}{dt}\frac{d\log\alpha}{dt}
-\frac12\left(\frac{d\log\alpha}{dt}\right)^2-\frac{d^2\log\alpha}{dt^2}\right].
\end{gather*}
The function $F$ can therefore be obtained by integrating
\begin{gather*}
v\frac{\partial F}{\partial x}+\left(\frac{3}{2}\frac{v^2}{x}+\frac{a_0(0)^2}{2}x^3+2\lambda x\right)
\frac{\partial F}{\partial v}=0.
\end{gather*}
In employing the {\it characteristics method}~\cite{CL09SRicc}, we f\/ind that $F$ must be constant along the integral
curves of the so-called {\it characteristic system}, namely
\begin{gather*}
\frac{dx}v=\frac{dv}{\frac{3}{2}\frac{v^2}{x}+\frac{a_0(0)^2}{2}x^3+2\lambda x}.
\end{gather*}
Let us focus on the region with $v\neq0$, i.e.\
$\mathcal{O}\equiv\{(x,v)\in{\rm T}\mathbb{R}_0\,|\,v\neq0\}$.
We obtain from the previous equations that
\begin{gather*}
\frac{dv}{dx}=\frac{3}{2}\frac{v}{x}+\frac{a_0(0)^2}{2}\frac{x^3}{v}+2\lambda\frac xv.
\end{gather*}
Let us focus on the case $x>0$; the other case can be obtained in a~similar way and leads to the same result.
Multiplying on right and left by $v/x$ and def\/ining $w\equiv v^2$ and $z\equiv x^2$, we obtain
\begin{gather*}
\frac{dw}{dz}=\frac{3}{2}\frac{w}{z}+\frac{a_0(0)^2}{2}z+2\lambda.
\end{gather*}
As this equation is linear, its general solution can be easily derived to obtain
\begin{gather*}
w(z)=\big(a_0(0)^2z^{1/2}-4\lambda z^{-1/2}+\xi\big)z^{3/2},
\end{gather*}
for an arbitrary real constant $\xi$.
From here, it results
\begin{gather*}
-a_0(0)^2x+\frac{v^2}{x^3}+\frac{4\lambda}{x}=\xi.
\end{gather*}
Consequently, $F$ is any function of the form $F=F(\xi)$, for instance,
\begin{gather*}
F=-a_0(0)^2x+\frac{v^2}{x^3}+\frac{4\lambda}x,
\qquad
(x,v)\in\mathcal{O}.
\end{gather*}
In principle, $F$ was def\/ined only on $\mathcal{O}$.
Nevertheless, as this region is an open and dense subset of ${\rm T}\mathbb{R}_0$, and in view of the expression for
$F$, we can extend it dif\/ferentiably to ${\rm T}\mathbb{R}_0$ in a~unique way.
Since $F$ is a~constant of motion on $\mathcal{O}$, it trivially becomes so on the whole ${\rm T}\mathbb{R}_0$.

Summarising, we have proved the following.
\begin{theorem}
\label{Th:IntGam}
A second-order Gambier equation
\begin{gather}
\label{QLSchemeGambier}
\frac{d^2x}{dt^2}=
\frac{3}{2x}\left(\frac{dx}{dt}\right)^2+\frac{1}{a_0}\frac{d{a}_0}{dt}\frac{dx}{dt}+\frac{a_0^2}{2}x^3-2a_2x,
\end{gather}
admits a~constant of motion of the form
\begin{gather}
\label{con}
F=-a_0(0)^2\bar{x}+\frac{\bar{v}^2}{\bar{x}^3}+\frac{4\lambda}{\bar{x}},
\end{gather}
where $\lambda$ is a~real constant, $\bar x=\alpha x$ and $\bar v=\delta dx/dt+\gamma x$, with $\alpha$ being
a~particular positive solution with $\alpha(0)=1$ of
\begin{gather}
\label{eqr}
\frac{d^2\log\alpha}{dt^2}=
\frac{2\lambda a_0^2}{a_0(0)^2\alpha^2}+2a_2+\frac{1}{a_0}\frac{da_0}{dt}\frac{d\log\alpha}{dt}-\frac12\left(\frac{d\log\alpha}{dt}\right)^2
\end{gather}
and $\gamma$, $\delta$ are determined from $\alpha$ by means of the conditions~\eqref{con6}.
\end{theorem}

The above proposition can be employed to derive a~constant of motion for certain families of second-order Gambier
equations.
For instance, if we start by a~Gambier equation~\eqref{QLSchemeGambier} with $a_2=-\lambda a_0^2/a_0(0)^2$, for
a~certain real constant $\lambda$, then $\alpha=1$ is a~solution of~\eqref{eqr}.
In view of~\eqref{con6}, $\gamma=0$ and $\delta=a_0(0)/a_0$.
Therefore, Theorem~\ref{Th:IntGam} establishes that the second-order Gambier equation~\eqref{QLSchemeGambier} admits
a~constant of motion
\begin{gather}
\label{EasyCons}
F=-a_0(0)^2x+\frac{a_0(0)^2}{a_0^2x^3}\left(\frac{dx}{dt}\right)^2+\frac{4\lambda}x.
\end{gather}

Consider now a~general second-order Gambier equation~\eqref{QLSchemeGambier} and let us search for a~constant of
motion~\eqref{con} with $\lambda=0$.
In this case,~\eqref{eqr} can be brought into a~Riccati equation
\begin{gather*}
\frac{dw}{dt}=2a_2+\frac{1}{a_0}\frac{da_0}{dt}w-\frac12w^2,
\end{gather*}
where $w\equiv d\log\alpha/dt$, whose solutions can be investigated through many methods~\cite{CarRamGra,CarRam}.
The derivation of a~particular solution provides a~constant of motion for the second-order Gambier
equation~\eqref{QLSchemeGambier} that can be obtained through the previous theorem.
Additionally, this particular solution can be used to obtain the general solution of the Riccati
equation~\cite{Dissertationes,CarRamGra}, which in turns could be used to derive new constants of motion for the
second-order Gambier equation~\eqref{QLSchemeGambier}.

Note that all the above procedure depends deeply on the fact that $\lambda$ is a~constant.
Recall that if $\lambda$ is not a~constant, then $g_\bigstar X$ does not admit any $t$-independent constant of motion.
Nevertheless, other methods can potentially be applied in this case.
For instance, using that $S(V_0,V_0)$ is a~quasi-Lie scheme, we can derive the group $\mathcal{G}(V_0)$ of this scheme
and to use an element $h\in\mathcal{G}(V_0)$ to transform $g_\bigstar X$ into other Lie system $h_\bigstar g_\bigstar
X$ of the same type, e.g.\
another of the form $h_\bigstar g_\bigstar X=\xi_2(t)(c_1X_1+c_2X_2+c_3X_3)$, with $c_1,c_2,c_3\in\mathbb{R}$, the
vector f\/ields $X_1$, $X_2$, $X_3$ are those of~\eqref{VFKS2}, and $\xi_2(t)$ is any $t$-dependent nonvanishing function.
As this system is, up to a~$t$-parametrization, a~$t$-independent vector f\/ield, it admits a~local $t$-independent
constant of motion.
By inverting the $t$-dependent changes of variables $h$ and $g$, it gives rise to a~$t$-dependent constant of motion of~$X$ and the corresponding second-order Gambier equation.

\section[Second-order Gambier and Milne-Pinney equations]{Second-order Gambier and Milne--Pinney equations}\label{Section7}

Consider the KS2 equation~\eqref{KS2three} with $x>0$ (we
can proceed analogously for the case $x<0$).
The change of variables $x=1/y^2$, with $y>0$, transforms it into a~Milne--Pinney equation~\cite{Pi50}
\begin{gather}
\label{Pi}
\frac{d^2y}{d\tau^2}=-\omega(t(\tau))y-\frac{a_0(0)^2}{4y^3}.
\end{gather}
These equations admit a~description in terms of a~Lie system related to a~Vessiot--Guldberg Lie algebra isomorphic to
$\mathfrak{sl}(2,\mathbb{R})$~\cite{SIGMA,PW}.
This was employed in several works to describe their general solutions in terms of particular solutions of the same or
others dif\/ferential equations, e.g.\
Riccati equations and $t$-dependent harmonic oscillators~\cite{Dissertationes}.

Previous results allow us to describe the general solution of~\eqref{GamQuasiLie} in terms of particular solutions of
Riccati equations or $t$-dependent frequency harmonic oscillators.
Indeed, in view of Corollary~\ref{Cor:trans}, these equations can be transformed into a~KS2 equation through
a~$t$-dependent change of variables $\bar x(t)=\alpha(t)x(t)$ and a~$t$-reparametrisation $d\tau=(\alpha/\delta)\,dt$.
In turn, $y=1/\sqrt{\bar x}$ transforms~\eqref{KS2three} into~\eqref{Pi}, whose general solution $y(t)$ can be written
as~\cite{SIGMA}
\begin{gather*}
y(\tau)=\sqrt{k_1z_1^2(\tau)+k_2z_2^2(\tau)+2C(k_1,k_2,W)z_1(\tau)z_2(\tau)},
\end{gather*}
where $C^2(k_1,k_2,W)=k_1k_2+a_0(0)^2/(4W^{2})$, the functions $z_1(\tau)$, $z_2(\tau)$ are two linearly independent
solutions of the system
\begin{gather*}
\frac{d^2z}{d\tau^2}=-\omega(t(\tau))z,
\end{gather*}
and $W$ is the Wronskian related to such solutions.
Inverting previous changes of variables, the general solution for any second-order Gambier
equation~\eqref{QLSchemeGambier} reads
\begin{gather*}
x(t)=
\alpha^{-1}\left[k_1z_1^2(\tau(t))+k_2z_2^2(\tau(t))\pm2\sqrt{k_1k_2+a_0(0)^2/(4W^{2})}z_1(\tau(t))z_2(\tau(t))\right]^{-1}.
\end{gather*}
Therefore, we have proved the following proposition:
\begin{proposition}
The general solution of a~second-order Gambier equation
\begin{gather}
\label{Gam}
\frac{d^2x}{dt^2}=
\frac{3}{2x}\left(\frac{dx}{dt}\right)^2+\frac{1}{a_0}\frac{d{a}_0}{dt}\frac{dx}{dt}+\frac{a_0^2}{2}x^3-2a_2x,
\end{gather}
can be brought into the form
\begin{gather*}
x(t)=
\alpha^{-1}\left[k_1z_1^2(\tau(t))
+k_2z_2^2(\tau(t))\pm2\sqrt{k_1k_2+a_0(0)^2/(4W^{2})}z_1(\tau(t))z_2(\tau(t))\right]^{-1},
\end{gather*}
where $z_1(\tau)$ and $z_2(\tau)$ are particular solutions of the $\tau$-dependent harmonic oscillator
\begin{gather*}
\frac{d^2z}{d\tau^2}=-\omega(t(\tau))z=
-\frac{\delta^2}{2\alpha^2}\left(2a_2+\frac{\gamma}{a_0\delta}\frac{da_0}{dt}-\frac{1}{2}\frac{\gamma^2}{\delta^2}
-\frac{1}{\delta}\frac{d\gamma}{dt}+\frac{\gamma}{\delta^2}\frac{d\delta}{dt}\right)z,
\end{gather*}
and $W=z_1\,dz_2/d\tau-z_2\,dz_1/d\tau$, with $d\tau=\alpha/\delta\;dt$ and $\alpha$, $\delta$ and $\gamma$ certain
$t$-dependent functions satisfying~\eqref{con6}.
\end{proposition}

Many other similar results can be obtained in an analogous manner.
For instance, the theo\-ry of Lie systems was also used in~\cite{SIGMA} to prove that the general solution of
a~Milne--Pinney equation~\eqref{Pi} can be written as
\begin{gather*}
y(\tau)=
\sqrt{\frac{[k_1(x_1(\tau)-x_2(\tau))-k_2(x_1(\tau)-x_3(\tau))]^2-a_0(0)^2[x_2(\tau)-x_3(\tau)]/4}
{(k_2-k_1)(x_2(\tau)-x_3(\tau))(x_2(\tau)-x_1(\tau))(x_1(\tau)-x_3(\tau))}},
\end{gather*}
where $x_1(\tau)$, $x_2(\tau)$ and $x_3(\tau)$ are three dif\/ferent particular solutions of the Riccati equation
\begin{gather*}
\frac{dx}{d\tau}=-\omega(t(\tau))-x^2.
\end{gather*}
Proceeding as before, we can describe the general solution of a~second-order Gambier equation~\eqref{Gam} in terms of
solutions of these Riccati equations.
In addition, by applying the theory of Lie systems~\cite{CarRamGra} to Milne--Pinney equations~\eqref{Pi}, we can
obtain many other results about subfamilies of second-order Gambier equations.
In addition, the relation between second-order Gambier equations and Milne--Pinney equations enables us to obtain
several other results in a~simple way.
\begin{proposition}
The second-order Gambier equation~\eqref{GamQuasiLie} with $2a_2=a_{0}^{2}>0$, i.e.\
\begin{gather}
\label{Gamsp}
\frac{d^2x}{dt^2}=
\frac{3}{2x}\left(\frac{dx}{dt}\right)^2+\frac{1}{a_0}\frac{d{a}_0}{dt}\frac{dx}{dt}+\frac{a_0^2}{2}x^3-a_{0}^{2}x,
\end{gather}
can be transformed into the integrable Milne--Pinney equation
\begin{gather}
\label{EMP}
\frac{d^2y}{d\tau^2}-\frac y2+\frac{1}{4y^3}=0
\end{gather}
under the transformation $x=1/y^2$ and a~$t$-reparametrisation $d\tau=a_0dt$.
\end{proposition}
\begin{note}
For second-order Gambier equations~\eqref{GamQuasiLie} with $a_{0}=0$, the transformations given in the above
proposition map these equations into harmonic oscillators.
\end{note}
\begin{proposition}
A constant of motion of~\eqref{EMP} is given by
\begin{gather*}
I_{MP}=\frac{1}{2}\left[\left(\frac{dy}{d\tau}\right)^2-\left(\frac{y^2}2+\frac{1}{4y^2}\right)\right].
\end{gather*}
Using $y=x^{-1/2}$ transformation, we obtain a~constant of motion for the special second-order Gambier
equation~\eqref{Gamsp} of the form
\begin{gather}
\label{Con2}
I_{2G}=\frac{1}{4}\frac{1}{x^3a_{0}^{2}}\left(\frac{dx}{dt}\right)^2-\left(\frac{1}{2x}+\frac{x}{4}\right).
\end{gather}
\end{proposition}

Thus, we say the special equation~\eqref{Gamsp}, which yields a~constant of motion, is an integrable deformation of the
Milne--Pinney equation.
\begin{note}
Recall that Theorem~\ref{Th:IntGam} can be applied to those equations~\eqref{GamQuasiLie} where $a_2=-\lambda
a_0^2/a_0^2(0)$ giving rise to the constant of motion~\eqref{EasyCons}.
Observe that~\eqref{Gamsp} is of this type with $\lambda=-a_0(0)^2/2$.
Then, the constant of motion~\eqref{EasyCons} is, up to a~multiplicative constant, the same as~\eqref{Con2}.
Despite this, the above illustrates how certain results can readily be obtained through Milne--Pinney equations.
\end{note}

\section{A second quasi-Lie schemes approach\\ to second-order Gambier equations}\label{Section8}

Apart from our f\/irst approach, we can
provide a~second quasi-Lie scheme to study second-order Gambier equations.
This one is motivated by the fact that some cases of these equations, namely
\begin{gather*}
\frac{d^2x}{dt^2}=-\left(3x\frac{dx}{dt}+x^3\right)+f(t)+g(t)x+h(t)\left(x^2+\frac{dx}{dt}\right),
\end{gather*}
for arbitrary $t$-dependent functions $f$, $g$ and $h$, form a~particular subclass of second-order Riccati equations that
are {\it SODE Lie systems}~\cite{CLS12,CGL11}.
That is, when written as f\/irst-order systems
\begin{gather}
\frac{dx}{dt}=v,
\nonumber
\\
\frac{dv}{dt}=-\left(3xv+x^3\right)+f(t)+g(t)x+h(t)\left(x^2+v\right),
\label{FirstSR}
\end{gather}
they become Lie systems.
Indeed, these systems possess a~Vessiot--Guldberg Lie algebra $V_0$ isomorphic to $\mathfrak{sl}(3,\mathbb{R})$ given
by~\cite{CL09SRicc,GL12}
\begin{gather*}
X_1=v\dfrac{\partial}{\partial x}-\big(3xv+x^3\big)\dfrac{\partial}{\partial v},\qquad X_2=\dfrac{\partial}{\partial v},
\qquad
X_3=-\dfrac{\partial}{\partial x}+3x\dfrac{\partial}{\partial v},\\ X_4=
x\dfrac{\partial}{\partial x}-2x^2\dfrac{\partial}{\partial v},\qquad
 X_5=\big(v+2x^2\big)\dfrac{\partial}{\partial x}-x\big(v+3x^2\big)\dfrac{\partial}{\partial v},\\
  X_6=
2x\big(v+x^2\big)\dfrac{\partial}{\partial x}+2\big(v^2-x^4\big)\dfrac{\partial}{\partial v},\qquad
X_7=\dfrac{\partial}{\partial x}-x\dfrac{\partial}{\partial v}, \qquad X_8=
2x\dfrac{\partial}{\partial x}+4v\dfrac{\partial}{\partial v}.
\end{gather*}
In terms of these vector f\/ields,~\eqref{FirstSR} is the associated system to the $t$-dependent vector f\/ield
\begin{gather*}
X_1+f(t)X_2+\frac12g(t)(X_3+X_7)+\frac{h(t)}4(-2X_4+X_8).
\end{gather*}
Therefore, it is natural to analyse which Gambier vector f\/ields $X$ can be transformed through a~quasi-Lie scheme
into one of these Lie systems.

Despite the interest of the previous idea, the quasi-Lie scheme provided in the previous section cannot be employed to
determine all quasi-Lie systems of this type.
The reason is that $V_0$ is not included in $V_G$ and, therefore, there exists no $g\in\mathcal{G}(W_G)$ such that
$g_\bigstar X\in V_0(C^\infty(\mathbb{R}))\subset V_G(C^\infty(\mathbb{R}))$.
To solve this drawback, we now determine a~new quasi-Lie scheme $S(W_G,V_G')$ such that the space~$V_G'$ contains
$V_G+V_0$.
This can be done by def\/ining~$V'_G$ as the $\mathbb{R}$-linear space generated by the elements of~$V_G$ and the
vector f\/ields
\begin{gather*}
Y_{12}=\frac{\partial}{\partial x},
\qquad
Y_{13}=x^2\frac{\partial}{\partial x},
\qquad
Y_{14}=xv\frac{\partial}{\partial x},
\\
Y_{15}=x^3\frac{\partial}{\partial x},
\qquad
Y_{16}=x^4\frac{\partial}{\partial v},
\qquad
Y_{17}=v^2\frac{\partial}{\partial v}.
\end{gather*}
Indeed, using Tables~\ref{Ta} and \ref{Ta2}, we readily obtain that $[W_G,V_G']\subset V'_G$.
Moreover, as we already know that $W_G$ is a~Lie algebra, $S(W_G,V'_G)$ becomes a~quasi-Lie scheme.
\begin{table}\centering
\caption{Lie brackets $[Y_i,Y_j]$ with $i=4,8,11$ and $j=12,\ldots,17$.}\label{Ta2}
\vspace{1mm}

\begin{tabular}
{ | c | c | c | c | c | c | c | c | c |}
\hline
$[\cdot,\cdot]$ & $Y_{12}$ & $Y_{13}$ & $Y_{14}$ & $Y_{15}$ & $Y_{16}$ & $Y_{17}$
\\
\hline
$Y_4$ & $0$ & $0$ & $Y_{14}$& $0$ & $-Y_{16}$ & $Y_{17}$
\\
\hline
$Y_{8}$ & $-Y_{9}$ & $-Y_{7}$& $Y_{13}-Y_3$ & $-Y_{6}$ &$0$ & $2Y_3$
\\
\hline
$Y_{11}$ & $-Y_{12}$ & $Y_{13}$ & $0$ & $2Y_{15}$ &$4Y_{16}$ & $0$
\\
\hline
\end{tabular}
\end{table}

Since we impose that $g_\bigstar X$, which is given by~\eqref{Decomp}, must be of the form~\eqref{FirstSR} up to
a~$t$-reparametrisation, we obtain
\begin{gather}
\label{transf2}
g_\bigstar X=\sum_{\alpha=1}^{11}\bar{b}_\alpha(t)Y_\alpha=
\xi(t)\left[(Y_1-3Y_3-Y_6)+f(t)Y_9+g(t)Y_8+h(t)(Y_7+Y_4)\right],
\end{gather}
for a~certain $t$-dependent non-vanishing function $\xi(t)$.
Taking into account that the vector f\/ields $Y_1,\ldots,Y_{11}$ are linearly independent over $\mathbb{R}$, we see that
\begin{gather*}
\bar{b}_2=0,
\qquad
\bar{b}_5=0,
\qquad
\bar{b}_{10}=0,
\qquad
\bar{b}_{11}=0.
\end{gather*}
Bearing in mind the form of the coef\/f\/icients $\bar b_\alpha$ given by~\eqref{NewCoef} and recalling that
$\alpha>0$, we see that $\bar{b}_2=0$ entails $n=1$.
Meanwhile, conditions $\bar{b}_5=0$ and $\bar{b}_{11}=0$ entail $\sigma=0$ and
\begin{gather}
\label{cond}
\frac{1}{\alpha}\frac{d\alpha}{dt}=\frac{\gamma}{\delta},
\end{gather}
respectively.
In turn, $\sigma=0$ also entails that $\bar{b}_9=\bar b_{10}=0$.
Moreover, from~\eqref{transf2} we also see that
\begin{gather*}
\bar{b}_3=-3\bar{b}_1,
\qquad
\bar{b}_6=-\bar{b}_1,
\qquad
\bar{b}_7=\bar{b}_4.
\end{gather*}
From these conditions and $n=1$, $\sigma=0$, we obtain
\begin{gather}
\label{three}
\frac{a_0}{\alpha}=-\frac{\alpha}{\delta},
\qquad
\frac{a_0^2\delta}{\alpha^3}=\frac{\alpha}{\delta},
\qquad
\frac{1}{\delta}\frac{d\delta}{dt}=\frac{2\gamma}{\delta}-\frac{1}{a_0}\frac{da_0}{dt}.
\end{gather}
Since $\delta,\alpha>0$, it can readily be seen that the second equation is an immediate consequence of the f\/irst
one, which in turn implies $a_0=-\alpha^2/\delta$.
Using that $\delta(0)=\alpha(0)=1$, we obtain $a_0(0)=-1$.
Previous results along with~\eqref{cond} ensure that the last condition in~\eqref{three} holds.
Hence, $\bar{b}_1$, $\bar b_4$ and~$\bar{b}_8$ become
\begin{gather}
\bar{b}_1=-\frac{a_0}{\alpha},
\qquad
\bar{b}_4=a_1-\frac1{a_0}\frac{da_0}{dt}+\frac{3}{\alpha}\frac{d\alpha}{dt},
\nonumber
\\
\bar{b}_8=
-\frac{a_2\alpha}{a_0}+\frac{a_1}{a_0}\frac{d\alpha}{dt}+\frac{2}{a_0\alpha}\left(\frac{d\alpha}{dt}\right)^2
-\frac{1}{a_0}\frac{d^2\alpha}{dt^2},
\label{coe}
\end{gather}
and the remaining coef\/f\/icients $\bar b_\alpha$ are either zero or can be obtained from them.
\begin{proposition}
\label{GamRic}
Every Gambier vector field $X$ is a~quasi-Lie system relative to the quasi-Lie scheme $S(W_G,V_G')$ and the Lie
algebra $V_0$ if and only if $n=1$, $a_0(0)=-1$ and $\sigma=0$.
In such a~case, every $t$-dependent transformation $g\in\mathcal{G}(W_G)$ given by
\begin{gather}
\bar x=\alpha x,
\qquad
\bar v=-\frac{\alpha}{a_0}\left(\frac{d\alpha}{dt}x+{\alpha}v\right),
\label{Change3}
\end{gather}
transforms $X$ into a~Lie system
\begin{gather}
\frac{d\bar x}{d\tau}=\bar v,
\nonumber
\\
\frac{d\bar v}{d\tau}=
-3\bar x\bar v-{\bar x}^3+\bar{b}_4\bar{b}_1^{-1}\bar x+\bar{b}_8\bar{b}_1^{-1}\big({\bar x}^2+\bar v\big),
\label{speRiccati}
\end{gather}
where $d\tau=-a_0dt/\alpha$ and $\bar{b}_1$, $\bar{b}_4$ and $\bar{b}_8$ are given by~\eqref{coe}.
\end{proposition}

\section{Exact solutions for several second-order Gambier equations}\label{Section9}

As a~result of Proposition~\ref{GamRic}, we can apply to~\eqref{speRiccati} the techniques of the theory of Lie systems
so as to devise, by inverting the change of variables~\eqref{Change3}, new properties of the Gambier equations related
to second-order Riccati equations.
For instance, we obtain new exact solutions of some of these families of second-order Gambier equations.

In the study of every Lie system, special relevance takes the algebraic structure of its Vessiot--Guldberg Lie algebras.
When a~Lie system possesses a~solvable Vessiot--Guldberg Lie algebra, we can apply several methods to explicitly
integrate it (cf.~\cite{Dissertationes,CLS12,CarRamGra}).
Otherwise, the general solution of the Lie system usually needs to be obtained through approximative
methods~\cite{Pi12} or expressed in terms of solutions of other Lie systems~\cite{SIGMA}.

System~\eqref{speRiccati} is a~special case of a~Lie system related to a~Vessiot--Guldberg Lie algebra $V_0$ isomorphic
to $\mathfrak{sl}(3,\mathbb{R})$, which is simple and therefore dif\/f\/icult to integrate explicitly.
Nevertheless, we can prove that~\eqref{speRiccati} is in general related to a~Vessiot--Guldberg Lie algebra that is
a~proper Lie subalgebra of $V_0$.
Moreover, we next prove that $X$ is related to a~solvable Vessiot--Guldberg Lie algebra in some particular cases that
can be explicitly integrated.

System~\eqref{speRiccati} describes the integral curves of the $t$-dependent vector f\/ield
\begin{gather*}
X_t=Z_1+\bar{b}_4\bar{b}_1^{-1}Z_2+\bar{b}_8\bar{b}_1^{-1}Z_3.
\end{gather*}
For generic $t$-dependent functions $\bar b_1$, $\bar b_4$ and $\bar b_8$, the above system admits a~Vessiot--Guldberg
Lie algebra spanned by $Z_1=X_1$, $Z_2=(X_3+X_7)/2$, $Z_3=(X_8-2X_4)/4$ and their successive Lie brackets, which generates
a~proper Lie subalgebra $V$ of $V_0$.
More specif\/ically, we f\/irst have
\begin{gather*}
\left[Z_1,Z_2\right]=Z_3-Z_4,
\qquad
\left[Z_1,Z_3\right]=-(Z_1+Z_5)/2,
\qquad
[Z_1,Z_5]=Z_6,
\end{gather*}
where $Z_4=X_4$, $Z_5=X_5$, $Z_6=X_6$.
This shows that the smallest Lie algebra containing~$Z_1$,~$Z_2$ and~$Z_3$ must include these vector f\/ields and~$Z_4$,~$Z_5$ and~$Z_6$.
Since these vector f\/ields additionally satisfy the following commutation relations{\samepage
\begin{gather*}
\left[Z_1,Z_4\right]=Z_5,
\qquad
\left[Z_3,Z_4\right]=0,
\qquad
\left[Z_1,Z_6\right]=0,
\qquad
\left[Z_2,Z_3\right]=Z_2,
\qquad
\left[Z_2,Z_4\right]=-Z_2,
\\
\left[Z_2,Z_5\right]=Z_4-Z_3,
\qquad\!
\left[Z_2,Z_6\right]=Z_5-Z_1,
\qquad\!
\left[Z_3,Z_5\right]=(Z_5+Z_1)/2,
\qquad\!
\left[Z_3,Z_6\right]=Z_6,
\\
\left[Z_4,Z_5\right]=-Z_1,
\qquad
\left[Z_4,Z_6\right]=0,
\qquad
\left[Z_5,Z_6\right]=0,
\end{gather*}}

\noindent
we see that they span a~six-dimensional proper Lie subalgebra~$V$ of~$V_0$.
Moreover, it is easy to show that $V\simeq\langle Z_1+Z_5,Z_3-Z_4,Z_2\rangle\oplus_S\langle
Z_6,Z_1-Z_5,Z_3+Z_4\rangle$, where $\langle Z_1+Z_5,Z_3-Z_4,Z_2\rangle\simeq\mathfrak{sl}(2,\mathbb{R})$, namely
\begin{gather*}
[Z_1+Z_5,Z_3-Z_4]=-2(Z_1+Z_5),
\\
[Z_2,Z_3-Z_4]=2Z_2,
\qquad
[Z_1+Z_5,Z_2]=2(Z_3-Z_4),
\end{gather*}
the linear space $\langle Z_6,Z_1-Z_5,Z_3+Z_4\rangle$ is a~solvable ideal of $V$ and $\oplus_S$ stands for the
semidirect sum of $\langle Z_1+Z_5,Z_3-Z_4,Z_2\rangle$ by $\langle Z_6,Z_1-Z_5,Z_3+Z_4\rangle$.

Hence, $X$ is related to a~non-solvable Vessiot--Guldberg Lie algebra and it is not known a~general method to
explicitly integrate $X$ for arbitrary $\bar b_1$, $\bar b_4$ and $\bar b_8$.
Nevertheless, we can focus on a~particular instance of these functions so that $X$ can be related to a~solvable
Vessiot--Guldberg Lie algebra~$V_1$.
For example, consider the case $\bar{b}_4=0$.
We then have
\begin{gather*}
V_1=\langle Z_1,Z_3,Z_1+Z_5,Z_6\rangle.
\end{gather*}
Note that the derived series of $V_1$ become zero, i.e.\
$\mathcal{D}^1\equiv[V_1,V_1]=\langle Z_1+Z_5,Z_6\rangle$,
$\mathcal{D}^2=0$.
In other words, $V_1$ is solvable and we can expect to integrate it through some method.
Let us do so through the so-called {\it mixed superposition rules}, i.e.\
a generalisation of superposition rules describing the general solution of a~Lie system in terms of several particular
solutions of other systems and a~set of constants~\cite{GL12}.
In our case, it is known (cf.~\cite[p.~194]{GL12}) that the general solution of~\eqref{speRiccati} can be written in the form
\begin{gather*}
\bar x(\tau)=
\frac{\lambda_1{v}_{y_1}(\tau)+\lambda_2{v}_{y_2}(\tau)+\lambda_3{v}_{y_3}(\tau)}
{\lambda_1y_1(\tau)+\lambda_2y_2(\tau)+\lambda_3y_3(\tau)},\!
\qquad
\bar v(\tau)=
\frac{d}{d\tau}\left(\frac{\lambda_1{v}_{y_1}(\tau)+\lambda_2{v}_{y_2}(\tau)
+\lambda_3{v}_{y_3}(\tau)}{\lambda_1y_1(t)+\lambda_2y_2(\tau)+\lambda_3y_3(\tau)}\right)\!,
\end{gather*}
where $(\lambda_1,\lambda_2,\lambda_3)\in\mathbb{R}^3-\{(0,0,0)\}$ and $(y_i(\tau),v_{y_i}(\tau),a_{y_i}(\tau))$, with
$i=1,2,3$, are linearly independent solutions of the linear Lie system
\begin{gather*}
\frac{dy}{d\tau}=v_y,
\qquad
\frac{dv_y}{d\tau}=a_y,
\qquad
\frac{da_y}{d\tau}=\bar{b}_8\bar{b}_1^{-1}a_y.
\end{gather*}
The above Lie system is related to a~Vessiot--Guldberg Lie algebra spanned by
\begin{gather*}
W_1=v_y\frac{\partial}{\partial y}+a_y\frac{\partial}{\partial v_y},
\qquad
W_2=a_y\frac{\partial}{\partial a_y},
\qquad
W_3=2a_y\frac{\partial}{\partial v_y},
\qquad
W_4=-2a_y\frac{\partial}{\partial y},
\end{gather*}
which close the same commutation relations as $Z_1$, $Z_3$, $Z_1+Z_5$ and $Z_6$.
Hence, this system is related to a~solvable Vessiot--Guldberg Lie algebra and it can easily be integrated:
\begin{gather*}
a_y=\exp\left(\int^\tau\bar{b}_8(\tau')\bar{b}^{-1}_1(\tau')d\tau'\right),
\qquad
v_y=\int^\tau a_y(\tau')d\tau',
\qquad
y=\int^\tau v_y(\tau')d\tau'.
\end{gather*}

Since we can assume
\begin{gather*}
\lambda_1y_1(\tau)+\lambda_2y_2(\tau)+\lambda_3y_3(\tau)=
\int^{\tau}\int^{\tau'}\exp\left(\int^{\tau''}\bar{b}_8(\tau''')\bar{b}^{-1}_1(\tau''')d\tau'''\right)d\tau''d\tau',
\end{gather*}
we obtain the solution of~\eqref{speRiccati} for $\bar{b}_4=0$, i.e.\
\begin{gather*}
\bar x(\tau)=
\frac{\int^{\tau}\exp\left(\int^{\tau'}\bar{b}_8(\tau'')\bar{b}^{-1}_1(\tau'')d\tau''\right)d\tau'}
{\int^{\tau}\int^{\tau'}
\exp\left(\int^{\tau''}\bar{b}_8(\tau''')\bar{b}^{-1}_1(\tau''')d\tau'''\right)d\tau''d\tau'},
\qquad
\bar v(\tau)=\frac{d\bar x}{d\tau}(\tau).
\end{gather*}

From above results, we have the following proposition.
\begin{proposition}
Given a~second-order Gambier equation with $n=1$, $a_0(0)=-1$, $\sigma=0$ and a~particular solution $\alpha(t)$ with
$\alpha(0)=1$ of
\begin{gather}
\label{EasyEqu}
a_1=\frac{1}{a_0}\frac{d a_0}{dt}-\frac{3}{\alpha}\frac{d\alpha}{dt},
\end{gather}
its general solution reads
\begin{gather*}
x(\tau(t))=
\frac{1}{\alpha(\tau(t))}
\frac{\int^{\tau(t)}\exp\left(\int^{\tau'}\bar{b}_8(\tau'')\bar{b}^{-1}_1(\tau'')d\tau''\right)d\tau'}
{\int^{\tau(t)}\int^{\tau'}\exp\left(\int^{\tau''}\bar{b}_8(\tau''')\bar{b}^{-1}_1(\tau''')d\tau'''\right)d\tau''d\tau'},
\end{gather*}
where $d\tau=-a_0dt/\alpha$ and $\bar{b}_1$ and $\bar b_8$ are given by~\eqref{coe}.
\end{proposition}
\begin{note}
Since equation~\eqref{EasyEqu} can be easily integrated, the above proposition shows that we can integrate exactly
every second-order Gambier equation with $n=1$, $a_0(0)=-1$ and $\sigma=0$.
\end{note}

\section{Conclusions}\label{Section10}

Two quasi-Lie schemes have been introduced to analyse the second-order Gambier equations.
Our f\/irst quasi-Lie scheme has been used to recover previous results concerning the reduction of such equations to
reduced canonical forms from a~geometric clarifying approach, which allowed us to f\/ill a~gap in the previous
literature.
This quasi-Lie scheme also led to determine some quasi-Lie systems related to certain second-order Gambier equations,
which enabled us to transform them into second-order Kummer--Schwarz equations.
We have expressed the general solutions of such second-order Gambier equations in terms of particular solutions of
$t$-dependent frequency harmonic oscillators and Riccati equations.
Additionally, new constants of motion were derived for some of them.

The introduction of a~second quasi-Lie scheme resulted in the description of an integrable family of second-order
Gambier equations related to second-order Riccati equations.

\subsection*{Acknowledgements}

The research of J.F.~Cari\~nena and J.~de Lucas was supported by the Polish National
Science Centre under the grant HARMONIA Nr 2012/04/M/ST1/00523.
They also acknowledge partial f\/inancial support by research projects MTM--2009--11154 (MEC) and E24/1 (DGA).
J.~de Lucas would like to thank for a~research grant FMI40/10 (DGA) to accomplish a~research stay in the University of
Zaragoza.

\pdfbookmark[1]{References}{ref}

\LastPageEnding

\end{document}